\documentclass{article}
\usepackage[utf8]{inputenc}
\usepackage{graphicx}
\usepackage{booktabs}
\usepackage{hyperref}
\usepackage{color, soul}

\title{Using convolutional neural networks for the classification of breast cancer images.}
\author{Eric Bonnet \\
Centre National de Recherche en Génomique Humaine (CNRGH)\\
CEA, Université Paris--Saclay, Evry, France\\
eric.bonnet@cea.fr}
\date{March 2024}

\begin{document}

\maketitle

\begin{abstract}
\textbf{Motivation:} Breast cancer is the leading cancer type in women worldwide. An important part of breast cancer staging is the assessment of the sentinel axillary node for early signs of tumor spreading. However, this assessment by pathologists is not always easy and retrospective surveys often requalify the status of a high proportion of sentinel nodes. Convolutional Neural Networks (CNNs) are a class of deep learning algorithms that have shown excellent performances in the most challenging visual classification tasks, with numerous applications in medical imaging. In this study I compare twelve different CNNs and different hardware acceleration devices for the detection of breast cancer from microscopic images of breast cancer tissue. Convolutional models are trained and tested on two public datasets. The first one is composed of more than 300,000 images of sentinel lymph node tissue from breast cancer patients, while the second one has more than 220,000 images from inductive breast carcinoma tissue, one of the most common forms of breast cancer. Four different hardware acceleration cards were used, with an off--the--shelf deep learning framework. The impact of transfer learning and hyperparameters fine-tuning are tested.\\

\textbf{Results:}  Hardware acceleration device performance can improve training time by a factor of five to twelve, depending on the model used. On the other hand, increasing convolutional depth will augment the training time by a factor of four to six times, depending on the acceleration device used. Increasing the depth and the complexity of the model generally improves performance, but the relationship is not linear and also depends on the architecture of the model. The performance of transfer learning is always worse compared to a complete retraining of the model. Fine-tuning the hyperparameters of the model improves the results, with the best model showing a performance comparable to state--of--the--art models. \\

\textbf{Availability:} All the models tested in this study, including the code for tuning some of the models, are available on GitHub (\url{https://github.com/erbon7/pcam\_analysis}).\\
\end{abstract}

\section*{Introduction}
Pathologists have been making diagnoses using digital images of glass microscope slides for many decades. In recent years, advances in slide scanning techniques has allowed the full digitization of microscopic stained tissue sections. There are many advantages to the digitalization of such images, including standardization, reproducibility, the ability to create  workflows, remote diagnostics, immediate access to archives and easier sharing among expert pathologists \cite{griffin2017digital}.
Breast cancer is the leading cancer type in women worldwide, with an estimated 2 million new cases and 627,000 deaths in 2018. Breast cancer staging refers to the process of describing the tumor growth or spread. Accurate staging by pathologists is an essential task that will determine the patient's treatment and his chances of recovery (prognosis). An important part of breast cancer staging is the assessment of the sentinel axillary node, a tissue commonly used for the detection of early signs of tumor spreading (metastasis). However, sentinel lymph nodes assessment by pathologists is not always easy and optimal. For instance, a retrospective survey performed in 2012 by expert pathologists requalified the status of a high proportion of sentinel nodes \cite{vestjens2012relevant}. 
Recently, deep learning algorithms have made major advances in solving problems that have resisted the machine learning and artificial intelligence community such as speech recognition, the activity of potential drug molecules, brain circuits reconstruction and the prediction of the effects of non-coding RNA mutation on gene expression and disease \cite{lecun2015deep}. Convolutional neural networks (CNNs) are a class of deep neural networks characterized by a shared-weight architecture of convolution kernels (or filters) that slide along input features and provide translation equivariant features known as feature maps. One of the main advantages of CNNs is that the network learns to optimize the filters through automated learning, requiring very little pre-processing compared to other machine learning techniques. Since their introduction in the 1990's \cite{lecun1989backpropagation}, CNNs have shown excellent performances in the most challenging visual classification tasks and are currently dominating this research field \cite{zeiler2014visualizing}. When applied to medical imaging, CNNs demonstrated excellent performance and have been successfully used for the identification of retinal diseases from fundus images \cite{ting2017development, kermany2018identifying, burlina2017automated}, tuberculosis from chest radiography images \cite{lakhani2017deep, ting2018clinical} and malignant melanoma from skin images \cite{esteva2017dermatologist}. CNNs have also been used for the detection of lymph node metastases in women with breast cancer in an algorithm competition known as CAMELYON16 (Cancer Metastases in Lymph Nodes Challenge), with the best models showing equal or slightly better performances than a panel of pathologists \cite{bejnordi2017diagnostic}. In this study, I use a dataset of more than 300,000 lymph node images derived from CAMELYON, known as the PCAM (Patch CAMELYON) dataset \cite{veeling2018rotation} and the IDC dataset, composed of more than 220,000 images derived from whole slide images of invasive ductal carcinoma tissue \cite{janowczyk2016deep,cruz2014automatic}, one of the most common forms of breast cancer. I used these datasets to characterize and analyze the performance of different CNNs network architectures and GPU accelerators, using a standard, off--the--shelf, deep learning computational library.

\section*{Material and methods}

\begin{figure*}[h]
\includegraphics[width=\textwidth]{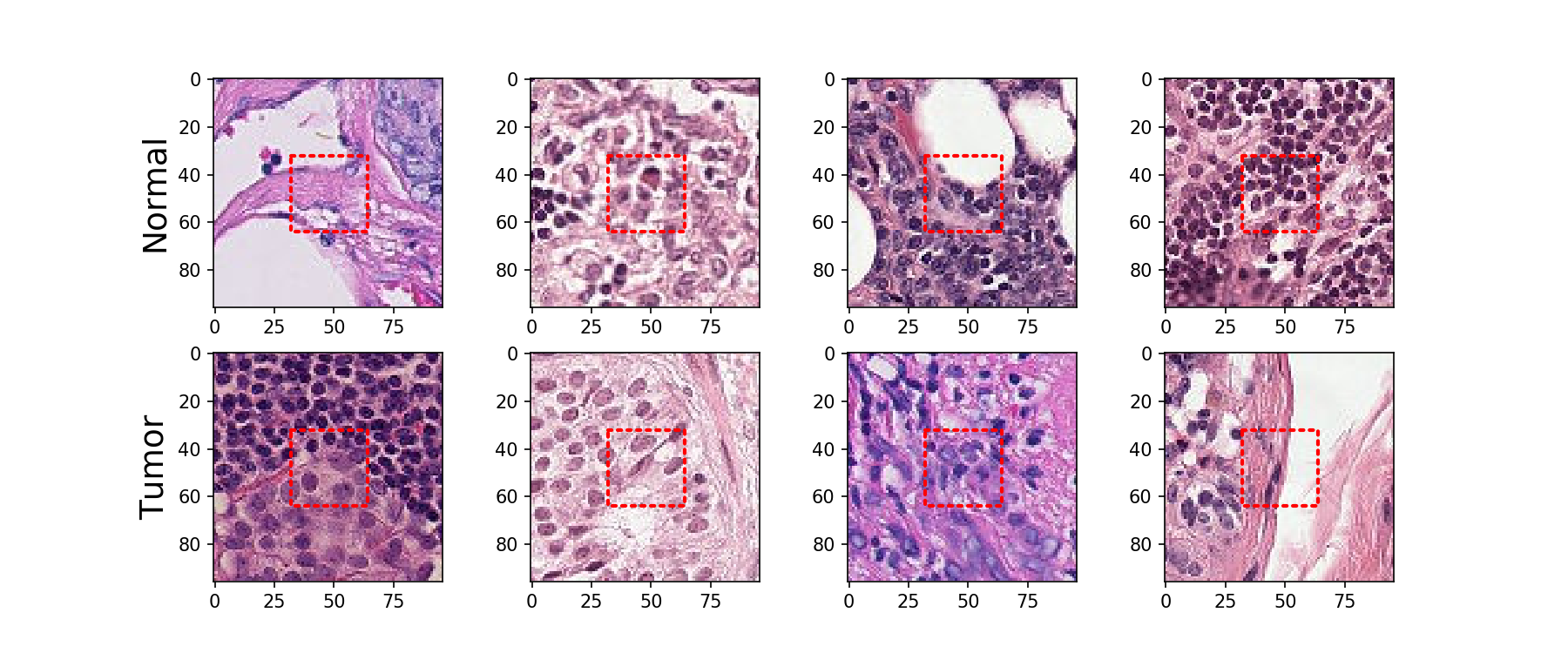}
\centering
\caption{Normal and tumor example images from the PCAM dataset. The red rectangle corresponds to the 32 x 32 pixels center. The presence of at least one pixel of tumor tissue in this region dictates a positive label (1), otherwise the image is labeled as negative(0).}
\label{fig:samples_pcam}
\end{figure*}

The PCAM dataset was downloaded from the original website (\url{https://github.com/basveeling/pcam}). All images have a size of 96 x 96 pixels, in three colors. The training set has 262,144 images (80 \% of the total), the validation set has 32,768 images (10 \%) and the test set also has 32,768 images (10 \%). All datasets have a 50/50 balance between positive (tumor present) and negative (tumor absent) samples. The patches of 96 x 96 pixels images were automatically extracted from the CAMELYON dataset \cite{veeling2018rotation}. For each image, a positive label indicates that the 32 x 32 pixel center of the image contains at least one pixel annotated as tumor tissue (Figure \ref{fig:samples_pcam}).

The IDC dataset was downloaded from 

\url{http://andrewjanowczyk.com/wp-static/IDC_regular_ps50_idx5.zip}. 

\begin{figure*}[h]
\includegraphics[width=.8\textwidth]{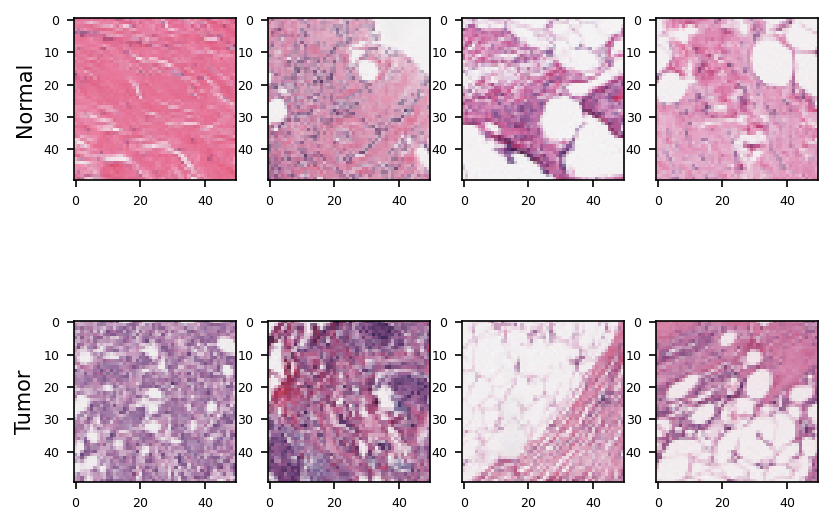}
\centering
\caption{Normal and tumor example images from the Inductive Ductal Carcinoma (IDC) dataset. Each image has a size of 50 x 50 pixels.}
\label{fig:samples_idc}
\end{figure*}

In total in this dataset there are 220,177 images of 50 x 50 pixels in three colors. Images classified as tumor by a pathologist are labelled as 1 while normal tissue images are labelled 0. The ratio between the two classes is 70 \% normal and 30 \% tumor for this dataset (Figure \ref{fig:samples_idc}).

All models were coded in Python (v3.7.5) using the TensorFlow (v2.4.1) and integrated Keras API (v2.4.0). Some additional statistics and performance indicators were calculated with the Sci-Kit Learn Python package \cite{pedregosa2011scikit}. The model performance was tested on four different GPU architectures: Nvidia Tesla K80 with 32 Gb of RAM, Nvidia Pascal P100 with 16 Gb of RAM, Nvidia Volta V100 with 32 Gb of RAM and Nvidia Ampere A100 with 80Gb of RAM. 

For each network architecture tested in this study, the same procedure is used: the model is trained on the training set for 15 epochs, with an evaluation on the validation set after each epoch. Depending on the accurracy value of the model, the weights are saved after each epoch to keep the best model, which is then evaluated on the test set. For some models, I used the KerasTuner framework with the Hyperband algorithm to optimize some hyperparameters \cite{omalley2019kerastuner}. 


\section*{Results}


Twelve different CNN models were used in this study, with different levels of depth and number of parameters (Table \ref{tab:models}). The models B2 and B6 were coded from scratch and have a relatively simple architecture. The B2 model has only two convolutional layers while the B6 model is slightly deeper and more complex with six convolutional layers. The architecture for the B6 model was inspired by the Kaggle models of F. Marazzi (\url{https://bit.ly/35wINGv}) and H. Mello (\url{https://bit.ly/3xwI6cl}). Note that the B6 model includes batch normalization \cite{ioffe2015batch} and dropout steps to improve performance. 

\begin{table}
 \caption{CNN models architecture and parameters.}
  \centering
  \begin{tabular}{lllll}
    \toprule
         Model name & Depth & Param. & Description or reference \\
    \hline\hline
    \addlinespace[2ex]
    B2 & 10 & 9.4 M& Basic model with two convolutional layers  \\
    B6 & 28 & 2.4 M & Basic model with six convolutional layers \\
    IV3 & 189 & 23.9M & Inception V3 \cite{szegedy2016rethinking} \\
    VGG19 & 25 & 143.7 M & VGG19 \cite{simonyan2014very} \\
    RN50 & 107 & 25.6 M & ResNet50 \cite{he2016deep} \\
    MN & 55 & 4.3 M & MobileNet \cite{howard2017mobilenets} \\
    DN121 & 242 & 8.1 M & DenseNet121 \cite{huang2017densely} \\
    DN169 & 338 & 14.3 M & DenseNet169 \cite{huang2017densely} \\
    DN201 & 402 & 20.2 M & DenseNet201 \cite{huang2017densely} \\
    ENB0 & 273 & 7.2 M & EfficientNet V2 B0 \cite{tan2021efficientnetv2} \\
    ENB1 & 337 & 8.2 M & EfficientNet V2 B1 \cite{tan2021efficientnetv2} \\
    ENB2 & 252 & 10.2 M & EfficientNet V2 B2 \cite{tan2021efficientnetv2} \\
    \addlinespace[2ex]
    \hline
  \end{tabular}
  \label{tab:models}
\end{table}

All the other models are CNN models that are part of the TensorFlow Keras library (Table \ref{tab:models}). They were developed and tested by several research groups on the Imagenet Challenge, a competition with hundreds of object categories and millions of images \cite{russakovsky2015imagenet}. For instance, InceptionV3 is a model created in 2015 with a very deep architecture (94 convolutional layers) that performs very well on various computer vision tasks \cite{szegedy2016rethinking}. As for most of the models available in the Keras llibrary, it is possible to load the model pre-weighted with ImageNet training weights, thus enabling transfer learning (TF). TF is a popular approach in deep learning where pre-trained models are used as the starting point on computer vision and natural language processing tasks in order to save computing and time resources. In this study I have used models both pre-trained with imagenet weights and fully re-trained with the two datasets. For the pre-trained version, only the last layers of the model are re-trained with the dataset (global average pooling layer, dense layer and final output). Of course, given that the number of training parameters is much greater in the case of the fully re-trained model, the computation time needed for training the model is also expected to be much longer. Table \ref{tab:models} is detailing the architecture and parameters for each of the models used in this study. Note that some CNN models could not be used with the IDC dataset because the images are smaller than the minimum size required by these models.

The computational running time was analysed for the for B2, B6 and the more complex InceptionV3 (IV3) model, both fully re-trained (F) and with transfer learning (TL) on the PCAM dataset. The results are shown in Table \ref{tab:time}. Note that the time corresponds to the average time observed for one epoch. We can compare the model architecture and the hardware GPUs acceleration effects. As expected, the running time is increasing with the complexity and depth of the model. The IV3-F model takes 4 to 10 times longer to train than the simple 2 convolutional layers B2 model, depending on the GPU card utilised. The B6 CNN model is taking 1.7 to 2 times longer than the B2 model to train. With the InceptionV3 model, using transfer learning is obviously saving a lot of training time, as a full model training is taking $\sim$3 times longer to train on all GPU models. In fact, even though the IVF-TL model (transfer learning) is much more complex, the running time is comparable to the B2 and B6 models. Regarding the different GPU cards tested here, more recent and powerful GPU cards decrease the computing time quite drastically, with an acceleration factor between 5 and 12 times for the most recent architecture tested here (A100) on all the CNN models compared to the oldest model tested here (K80). It is worth noting that the deepest model tested here can be fully trained in about one hour with a V100 or A100 GPU card. 

\begin{table}
 \caption{Run time in seconds for one epoch on different GPU architectures. NbCU: number of CUDA cores. Pp: processing power in GFlops. TL: transfer learning. F: full retraining.}
  \centering
  \begin{tabular}{lllllll}
    \toprule
        GPU card & NbCU & Pp & B2 & B6 & IV3-TL & IV3-F \\
    \midrule
    K80 (2014) & 4992 & 2400 & 325 & 562 & 468 & 1333 \\
    P100 (2016) & 3584 & 5000 & 70 & 121 & 143 & 393 \\
    V100 (2017) & 5120 & 7000 & 47 & 78 & 98 & 279 \\
    A100 (2020) & 6912 & 8000 & 26 & 57 & 100 & 270 \\
    
    \hline\hline
  \end{tabular}
  \label{tab:time}
\end{table}

\begin{figure}[h]
\includegraphics[width=\textwidth]{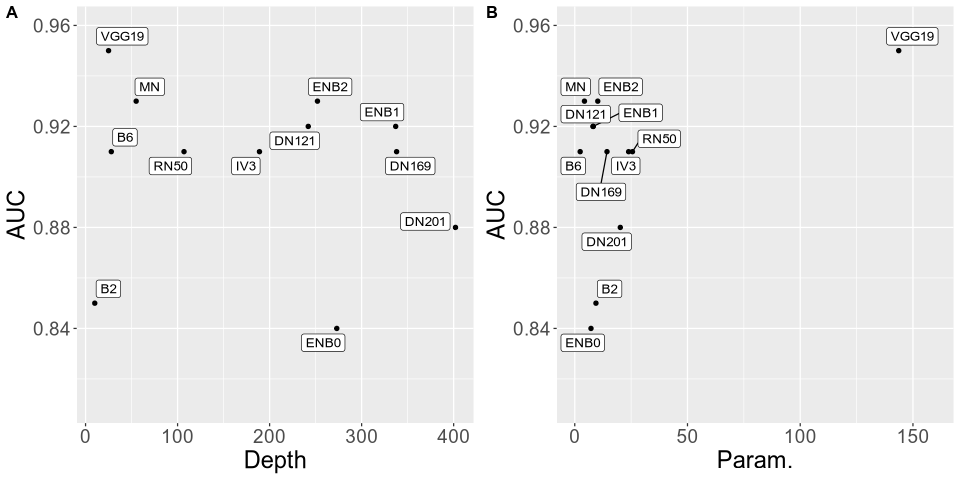}
\centering
\caption{Dotplots showing the AUC score versus the depth of the model, i.e. the total number of layers in the model (A) or the total number of parameters of the model (B) for the PCAM dataset.}
\label{fig:dotplot_auc}
\end{figure}

The performance of all the models on the PCAM and IDC datasets is described in Table \ref{tab:roc_pcam} and \ref{tab:roc_idc}. All the indicators are measured on the test sets. Most of the model show a very good performance, with AUC scores around 0.90 or above. However, when we look at the details, there are clear differences. For instance the AUC of the simple 2-layers B2 model is 0.85, increasing to 0.91 with the B6 model, which is slightly more complex. If we take the best performing models in terms of AUC score, for the PCAM dataset we have VGG19 (0.95), followed by the MobileNet (MN, 0.93) and the EfficientNet V2 B2 (ENB2, 0.93). For the IDC dataset, the best performing models are again VGG19 (0.95), together with MobileNet (MN, 0.95), and followed by the B6 (0.94), DenseNet 121 (DN121, 0.94) and EfficientNet V2 B2 (ENB2, 0.94). Looking at the AUC values versus depth and number of parameters of the models for the PCAM dataset (Figure \ref{fig:dotplot_auc}), we can see that the VGG19 model has a high AUC value but a low depth, while the ENB0 model has a very high depth but a very low AUC. For the number of parameters, VGG19 has a very high number of parameters and the best AUC, but models with a much lower number of parameters, such as MN or ENB2, also have high AUC values, in fact very close to the value for VGG19. Taken together, these results show that the more complex models perform better than very simple models (such as the B2 model), but the relationship is not entirely straightforward.

\begin{table}
 \caption{Performance of the models on the PCAM test set. AUC: area under the ROC curve.}
  \centering
  \begin{tabular}{lccc}
    \addlinespace[2ex]
    \toprule
        CNN model & Loss & Accuracy & AUC \\
    \hline\hline
    \addlinespace[1ex]
    \textit{Transfert learning} & & & \\
    \hline
    \addlinespace[1ex]
    IV3 & 0.46 & 0.78 & 0.88 \\
    VGG19 & 0.42 & 0.80 & 0.89 \\
    RN50 & 0.52 & 0.74 & 0.82 \\
    MN & 0.46 & 0.79 & 0.88 \\
    DN121 & 0.38 & 0.82 & 0.91 \\
    DN169 & 0.49 & 0.79 & 0.90 \\
    DN201 & 0.39 & 0.82 & 0.91 \\
    ENB0 & 0.53 & 0.73 & 0.81 \\
    ENB1 & 0.52 & 0.74 & 0.82 \\
    ENB2 & 0.54 & 0.73 & 0.80 \\

    \addlinespace[1ex]
    \textit{Full training} & & & \\
    \hline
    \addlinespace[1ex]
    B2 & 0.47 & 0.78 & 0.85 \\
    B6 & 0.44 & 0.84 & 0.91 \\    
    IV3 & 0.42 & 0.84 & 0.91 \\
    VGG19 & 0.43 & 0.86 & 0.95 \\
    RN50 & 0.57 & 0.82 & 0.91 \\
    MN & 0.53 & 0.84 & 0.93 \\
    DN121 & 0.76 & 0.82 & 0.92 \\
    DN169 & 0.46 & 0.83 & 0.91 \\
    DN201 & 0.75 & 0.80 & 0.88 \\
    ENB0 & 0.99 & 0.70 & 0.84 \\
    ENB1 & 0.89 & 0.80 & 0.92 \\
    ENB2 & 0.94 & 0.84 & 0.93 \\
    
    \addlinespace[2ex]
    \hline
  \end{tabular}
  \label{tab:roc_pcam}
\end{table}

\begin{table}
 \caption{Performance of the models on the invasive ductal carcinoma (IDC) breast cancer test set. AUC: area under the ROC curve.}
  \centering
  \begin{tabular}{lccc}
    \addlinespace[2ex]
    \toprule
        CNN model & Loss & Accuracy & AUC \\
    \hline\hline
    \addlinespace[1ex]
    \textit{Transfert learning} & & & \\
    \hline
    \addlinespace[1ex]
    VGG19 &  0.38 & 0.83 & 0.88 \\
    RN50 &  0.46 & 0.79 & 0.82 \\
    MN &  0.56 & 0.71 & 0.66 \\
    DN121 &  0.33 & 0.85 & 0.91 \\
    DN169 &  0.35 & 0.85 & 0.90 \\
    DN201 &  0.34 & 0.82 & 0.91 \\
    ENB0 &  0.48 & 0.78 & 0.81\\
    ENB1 & 0.49 & 0.79 & 0.81 \\
    ENB2 &  0.52 & 0.75 & 0.75\\

    \addlinespace[1ex]
    \textit{Full training} & & & \\
    \hline
    \addlinespace[1ex]
    B2 & 0.31 & 0.86 & 0.92 \\
    B6 &  0.28 & 0.87 & 0.94\\
    VGG19 & 0.25 & 0.90 & 0.95 \\
    RN50 & 0.28 & 0.88 & 0.93 \\
    MN &  0.26 & 0.89 & 0.95\\
    DN121 &  0.27 & 0.88 & 0.94 \\
    ENB2 & 0.28 & 0.88 & 0.94 \\
    DN201 &  0.31 & 0.88 & 0.93\\
    ENB0 &  0.35 & 0.86 & 0.92\\
    ENB1 &  0.43 & 0.84 & 0.91\\
    DN169 &  0.35 & 0.85 & 0.90 \\
    
    \addlinespace[2ex]
    \hline
  \end{tabular}
  \label{tab:roc_idc}
\end{table}

\begin{figure}[h]
\includegraphics[width=\textwidth]{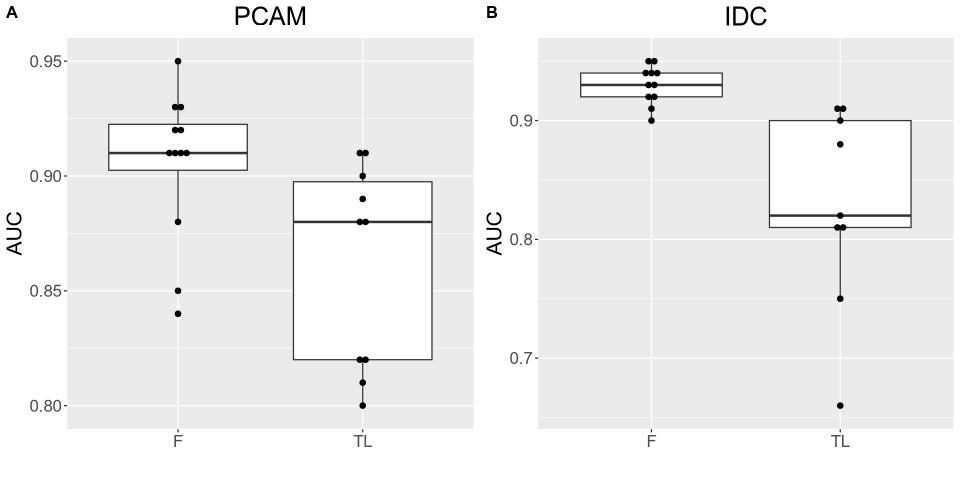}
\centering
\caption{Boxplots showing the AUC score for different CNN models for fully re-trained models (F) or with transfer learning (TL).}
\label{fig:boxplot_auc}
\end{figure}

It is worth noting that the best models are all models that were fully retrained on the data, not the models that used transfer learning. In fact, the models that are fully retrained have higher AUC values on average (Figure \ref{fig:boxplot_auc}) for both the PCAm and the IDC datasets, and if we look at the average of the AUC values there is a statistically significant difference (Welch Two Sample t-test, for the PCAM dataset t = -2.5527, df = 16.285, p-value = 0.02108, for the IDC dataset, t = -3.5998, df = 8.4843, p-value = 0.006335). The reason for this is probably that the visual structures (i.e. the filters) learned with the ImageNet dataset are not adapted to the PCAM images. Indeed, the object categories in the ImageNet dataset (such as "balloon", "strawberry", etc.) are very different from the types and structures seen in digital pathology images.

Fine tuning of the hyperparameters was done for the Inception V3 and the VGG19 models on the PCAM dataset. Two hyperparameters (Adam learning rate and batch size) were fine-tuned using the Keras Tuner with the hyperband algorithm. The performance is improved in the two models with an AUC of 0.95 for the Inception V3 model and an AUC of 0.96 for the VGG19 model. These performances are comparable to current state--of--the--art models for computational pathology analysis. It is within the top 5 best algorithms of the CAMELYON16 challenge \cite{bejnordi2017diagnostic} and is within the top 10 best models for the PCAM dataset (\url{https://tinyurl.com/3rhk6ph6}). The current best PCAM models have an AUC around 0.97 and implement rotation equivariant strategies \cite{cohen2016group, graham2020dense, weiler2018learning}. Indeed, histology images are typically symmetric under rotation, meaning that each orientation is equally as likely to appear. Rotation--equivariance removes the necessity lo learn this type of transformation from the data, thus allowing more discriminative features to be learned and also reducing the number of parameters of the model.

\section*{Conclusion}

Precise staging by expert pathologists of breast cancer axillary nodes, a tissue commonly used for the detection of early signs of tumor spreading, is an essential task that will determine the patient's treatment and his chances of recovery. However, it is a difficult task that was shown to be prone to misclassification. Algorithms, and in particular deep learning based convolutional neural networks, can help the experts in this task by analyzing fully digitized slides of microscopic stained tissue sections. In this study, I evaluated twelve different CNN architectures and different hardware acceleration devices for breast cancer classification on two different public datasets consisting of hundreds of thousands of images. The performance of hardware acceleration devices can improve the training time by a factor of five to twelve, depending on the model used. On the other hand, increasing the convolutional depth increases the training time by a factor of four to six, depending on the acceleration device used. More complex models tend to perform better than very simple ones, especially when fully retrained on the digital pathology dataset, but the relationship between model complexity and performance is not straightforward. Transfer learning from imagenet always performs worse than fully retraining the models. Fine-tuning the hyperparameters of the model improves the results, with the best model tested in this study showing very high performance, comparable to current state--of--the--art models.

\section*{Acknowledgements}

I wish to thank Claude Scarpelli and Jean-François Deleuze for their general support and discussions. I would also like to thank Christine Ménaché and Xavier Delaruelle for giving me access to the new FENIX infrastucture of the CEA cluster computer at the Très Grand Centre de Calcul (TGCC). Last, I would like to thank the TGCC support team for their patient and competent answers to my multiple questions and requests.


\end{document}